\newif\ifcheckpagelimits
\checkpagelimitstrue
\checkpagelimitsfalse

\ifcheckpagelimits

 \documentclass[nofootinbib,prl,aps,twocolumn,showpacs,showkeys,amsmath,amssymb,superscriptaddress,final,reprint,floatfix,longbibliography]{revtex4-1}

 \newcommand{\todo}[1]{}
\else
 \documentclass[prl,aps,twocolumn,showpacs,showkeys,%
 amsmath,amssymb,superscriptaddress,final,reprint,floatfix,longbibliography]{revtex4-1}
 \newcommand{\todo}[1]{{\pdfmargincomment[icon=Note,color=pink]{#1}}}
\fi

\usepackage{lineno}
\usepackage{mathptmx}
\usepackage{subfigure}
\usepackage[utf8]{inputenc}
\usepackage{dcolumn}
\usepackage{amsmath,amssymb}
\usepackage{bm}
\usepackage{color}
\usepackage{overpic}
\usepackage{latexsym}
\usepackage{epstopdf}
\usepackage[english]{babel}
\usepackage{latexsym}
\usepackage{stmaryrd}

\definecolor{mygrey}{gray}{0.35}
\definecolor{myblue}{rgb}{0.2,0.2,0.8}
\definecolor{myzard}{cmyk}{0,0,0.05,0}
\definecolor{mywhite}{rgb}{1,1,1}
\definecolor{myred}{rgb}{1,0.,0.3}

\usepackage[colorlinks=true,citecolor=myblue,linkcolor=myred,urlcolor=myblue]{hyperref}

 \def\ee{\mathord{\rm e}}

\renewcommand{\ee}{{\rm e}}
\def\beq{\begin{equation}}
\def\eeq{\end{equation}}

\usepackage{graphicx}
\usepackage{soul}
\usepackage{epstopdf}
\usepackage{dsfont}

\newcommand{\ket}[1]{\left\vert #1 \right\rangle}
\newcommand{\bra}[1]{\left\langle #1 \right\vert}
\newcommand{\ketbra}[2]{\ket{ #1}\bra{ #2} }
\newcommand{\bla}[1]{\left( #1 \right)}
\newcommand{\blb}[1]{\left[ #1 \right]}

\def \ket#1{|#1\rangle}
\def \bra#1{\langle#1|}

\def \be{\begin{equation}}
\def \ee{\end{equation}}
\def \ba{\begin{array}}
\def \ea{\end{array}}
\def \bea{\begin{eqnarray}}
\def \eea{\end{eqnarray}}

\renewcommand{\phi}{\varphi}

\usepackage{tabularx,ragged2e,booktabs}
\newcolumntype{C}[1]{>{\Centering}m{#1}}

\begin{document}

\title{Adiabatic quantum parameter amplification for generic robust quantum sensing}
\author{Yu Liu}
\affiliation{School of Physics \& Center for Quantum Optical Science, Huazhong University of Science and Technology, Wuhan 430074, P. R. China}

\author{Zijun Shu}
\affiliation{School of Physics \& Center for Quantum Optical Science, Huazhong University of Science and Technology, Wuhan 430074, P. R. China}

\author{Martin B. Plenio}
\affiliation{Institut f\"{u}r Theoretische Physik \& IQST, Albert-Einstein Allee 11, Universit\"{a}t Ulm, 89069 Ulm, Germany}

\author{Jianming Cai}
\email{jianmingcai@hust.edu.cn}
\affiliation{School of Physics \& Center for Quantum Optical Science, Huazhong University of Science and Technology, Wuhan 430074, P. R. China}

\pacs{03.67.Ac, 37.10.Vz, 75.10.Pq}

\begin{abstract}
Quantum enhanced sensing provides a powerful tool for the precise measurement of physical parameters that is applicable in many areas of science and technology. The achievable gain in sensitivity is largely limited by the influence of noise and decoherence. Here, we propose a paradigm of adiabatic quantum parameter amplification to overcome this limitation. We demonstrate that it allows to achieve generic robust quantum sensing, namely it is robust against noise that may even acting on the same degree of freedom as the field. Furthermore, the proposal achieves entanglement-free Heisenberg limit sensitivity that surpasses the limit of classical statistics.
\end{abstract}

\date{\today}

\maketitle

{\it Introduction.---} The emergent fields of quantum information science and quantum technologies
are promising devices that make use of quantum properties to achieve performances that exceed
what is possible in the realm of classical physics and promise considerable impact in physics, material science and biology. In this context, quantum sensing and metrology has attracted increasing interest because a relatively modest number of quantum systems under experimental control may already achieve a very
useful enhancement of performance \cite{Gio2011}. Indeed, by allowing increasingly precise measurement of physical
parameters and sensitive detection, this may enable new potential applications in a wide range of different
areas. A great deal of effort has been invested over the last two decades to develop high precision
quantum sensing and metrology protocols \cite{Wineland1992,Choi08,Roy08,Napo11,Boi07,Gold11,Cap12}, to study their limitations imposed by environmental  noise \cite{Esc11,Huel97,Boixo2008,Kok12,chin2012quantum,Smir16,Dur16} and to develop means to mitigate the
deleterious effects of noise \cite{Preskill2000,Macchiavello2002,Chaves13,Kotler11,Dur14,Kess14,Arad14,Lu14,Kotler14,Bau16,unden2016quantum,Dorner12,Genes13,Sza14,Sun16}.

A typical procedure of quantum metrology, e.g., the Ramsey method, uses a quantum particle that is subjected 
to a time evolution under a Hamiltonian that depends on a parameter that we wish to determine. This involves a sequence of interrogation cycles  with the interrogation time duration $T$. As the measurement sensitivity scales as $1/\sqrt{T}$, it is beneficial to extend $T$ to its maximally possible value. This value is limited by the presence of noise. Alternatively, the sensitivity of metrology may be enhanced using N quantum probes that are prepared in a multipartite entangled states and then each subjected to the same time evolution \cite{Wineland1992,Gio2011,Boi07} thereby achieving an $N$ times more rapid accumulation of phase. Nevertheless, quantum entanglement particularly in maximally entangled state is quite fragile under environmental noise. It has been shown that noise would degrade or even completely eliminate the improvement in the scaling of precision \cite{Huel97}, and thus hinder the implementation of Heisenberg limited quantum sensing. The application of active methods such as dynamical decoupling and quantum error correction in order to improve the sensitivity of quantum metrology against noise \cite{Macchiavello2002,Kotler11,Chaves13,Dur14,Kess14,Arad14,Lu14,Kotler14,unden2016quantum,Sun16,Bau16} aiming to reinstate the Heisenberg limit has received considerable attention over the years. Nevertheless, both dynamical decoupling and quantum error correction have limitations of their own. Dynamical decoupling based techniques work efficiently only for oscillating fields, while quantum error correction can only improve measurement sensitivity only under limited types of noise \cite{Sekatski2016}. Most importantly, it remains an open question how to achieve robust quantum parameter estimation of local Hamiltonians in the presence of parallel phase noise \cite{Macchiavello2002,Dur14}, e.g. for local Hamiltonian $H_s=b \sigma_z$ with noise $H_{noise}=\delta(t) \sigma_z $.

In this work, we propose a paradigm of adiabatic quantum metrology which consists of sensing and probe systems to determine a local parameter e.g. $H_s=b \sigma_z$. The sensing systems are adiabatically prepared into the ground state of a parameter-dependent local driven Hamiltonian. The energy gap protection makes it robust against noise that may even be acting along the $\hat{z}$-direction. By engineering an effective state-dependent interaction between the sensing systems and a probe system, the ground state encoded parameter information can be extracted. The interrogation time is mainly limited by the coherence time of the probe system which itself does not need to interact with the field that is to be measured. Hence it can be assumed to be unaffected by noise, e.g. originating from nuclear spin environments, may thus have a significantly longer coherence time. We demonstrate that it is possible to achieve Heisenberg limit scaling without involving entanglement as a resource in the sensing systems, and provide a new perspective concerning the role of entanglement in connection with quantum metrology \cite{Caves08,Cze15,Aug16}. The present idea of adiabatic quantum metrology is readily implementable in current state-of-art experiment, e.g. trapped ion and superconducting qubits, and may find application in a broad range of scenarios, ranging from the measurement of magnetic field and electric fields to that of forces.

{\it Adiabatic quantum parameter amplification.---} We start from $N$ two-level quantum sensing systems whose eigenstates are denoted as $\{\ket{0,1}\}$. The $k$-th sensing system, in an interaction picture with $\omega_0 \sigma_z^{(k)}$ is governed by the following Hamiltonian as
\begin{equation}
    H_k(t) = b \sigma_z^{(k)} + \lambda_k(t)\sigma_x^{(k)}, \label{eq:local_ham}
\end{equation}
where $\sigma_z^{(k)}=\ketbra{0_k}{0_k} - \ketbra{1_k}{1_k}$, $\sigma_x^{(k)} = \ketbra{0_k}{1_k} + \ketbra{1_k}{0_k}$
are the corresponding Pauli operators, $\lambda_k(t)$ quantifies the strength of a time-dependent field (e.g. a
laser acting on an atom/ion or a microwave driving field on a spin) applied to the k-th system via $\sigma_x^{(k)}$, and $b$ is the physical parameter that we would like to estimate. We remark that the parameter $b$ may also arise from the interaction with a magnetic dipole via e.g. dipole-dipole interaction, and thus the present idea can be extended to the detection of single atomic or nuclear dipole moments \cite{Kotler14}. The above setting arises naturally in the presence of non-Markovian environments as $\lambda_k(t)\sigma_x^{(k)}$ corresponds to applying a continuous drive to protect a qubit against noise \cite{Cai12,Tim11}. It has be demonstrated that the lifetime of the eigenstates of the above Hamiltonian that we will exploit in our scheme can be dramatically prolonged \cite{Cai12}.

\begin{figure}[t]
\begin{center}
\includegraphics[width=8.5cm]{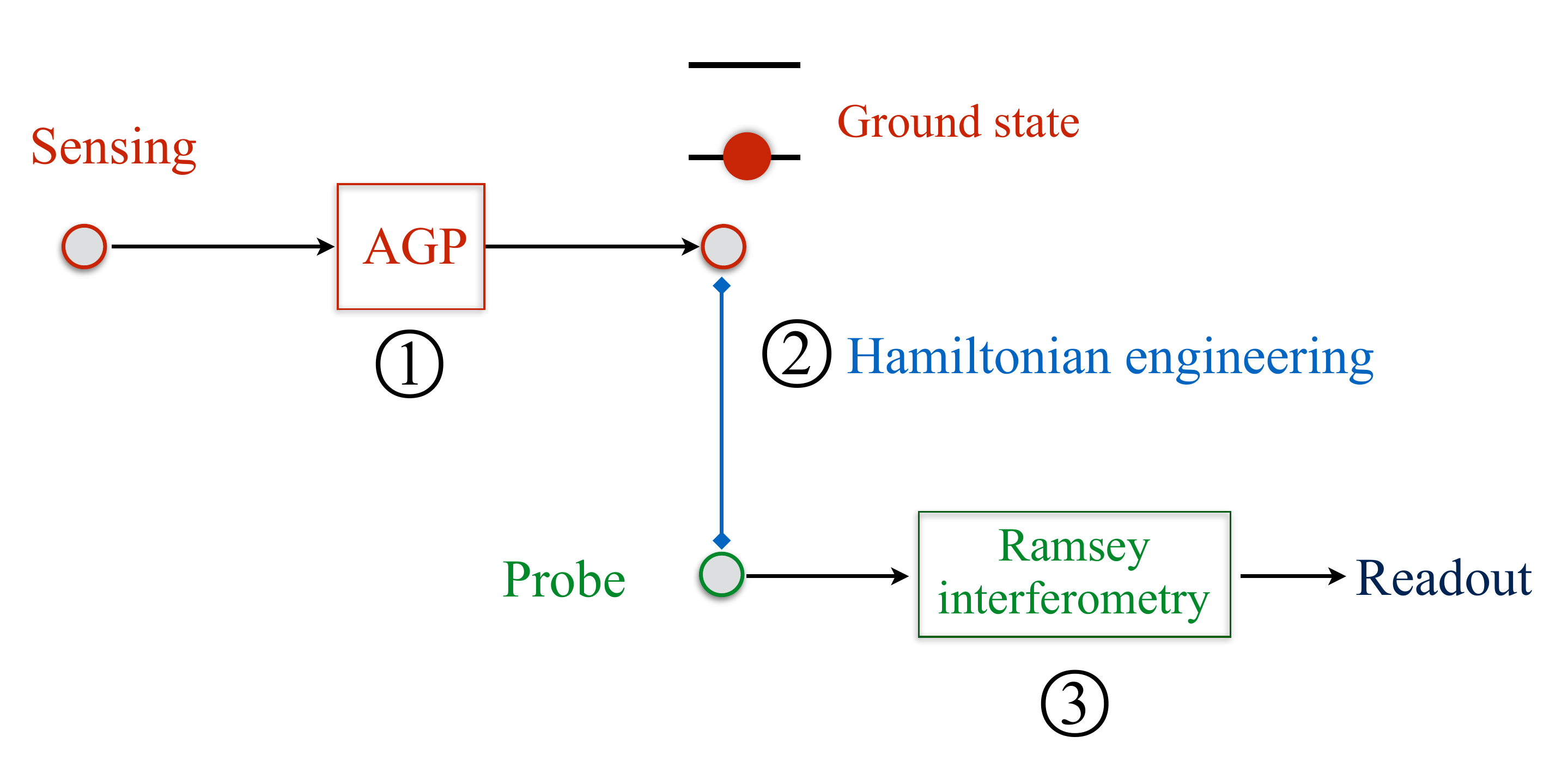}
\end{center}
\caption{(Color online) Schematic diabatic quantum parameter estimation. The sensing system is adiabatically prepared into the ground state (AGP) of a local Hamiltonian which contains the interaction
with a field quantified by the parameter $b$. The probe system couples with the sensing system via parameter-independent interactions that serves to amplify the field  by engineering the interaction to be state dependent, and to extract the parameter information following a Ramsay interferometry procedure.}
\label{fig:model}
\end{figure}

Instead of preparing the $N$ systems into a specific state and then applying the parameter-dependent
Hamiltonian followed by a final measurement, we encode the parameter information into the
ground state of $N$ independent systems, then amplify and transfer the parameter information
to the auxiliary probe system, see Fig.\ref{fig:model}. To this end, we chose the initial
value of the time-dependent parameter $\lambda_k(t)$ as $\lambda_k(0)\gg b$, and prepare the $k$-th
system in the state $\ket{\psi_k(0)}\equiv \ket{{\downarrow}_x^{(k)}} = \sqrt{\frac{1}{2}}\bla{\ket{0_k}-\ket{1_k}}$
that very closely approximates the ground state of the initial Hamiltonian $H_k(0)$. By slowly
decreasing the field $\lambda_k(t)$ until $\lambda_k(T_a)=h$, we adiabatically prepare the system to
the ground state of the Hamiltonian $H_k(T_a)=b\sigma_z^{(k)}+h\sigma_x^{(k)}$, namely
\begin{equation}
    \ket{\psi_k(T_a)} = \ket{G_{k}(b,h)}\equiv -\sin(\frac{\theta}{2})\ket{0_k}+\cos(\frac{\theta}{2})\ket{1_k},
\end{equation}
where $\cos\theta={b/\omega(b,h)}$ and $\sin\theta={h/\omega(b,h)}$ with $\omega(b,h)=\bla{h^2+b^2}^{1/2}$.
The ground state $\ket{G_k(b,h)}$ encodes the information on the value of the parameter $b$ albeit with a
sensitivity that is reduced due to the additional field $h$ as compared with the standard Ramsey interferometry method.
Moreover, the $N$ sensing systems are in a separable state, and thus independent measurements can only achieve standard quantum limit.

\begin{figure}[t]
\begin{center}
\hspace{0cm}
\includegraphics[width=8cm]{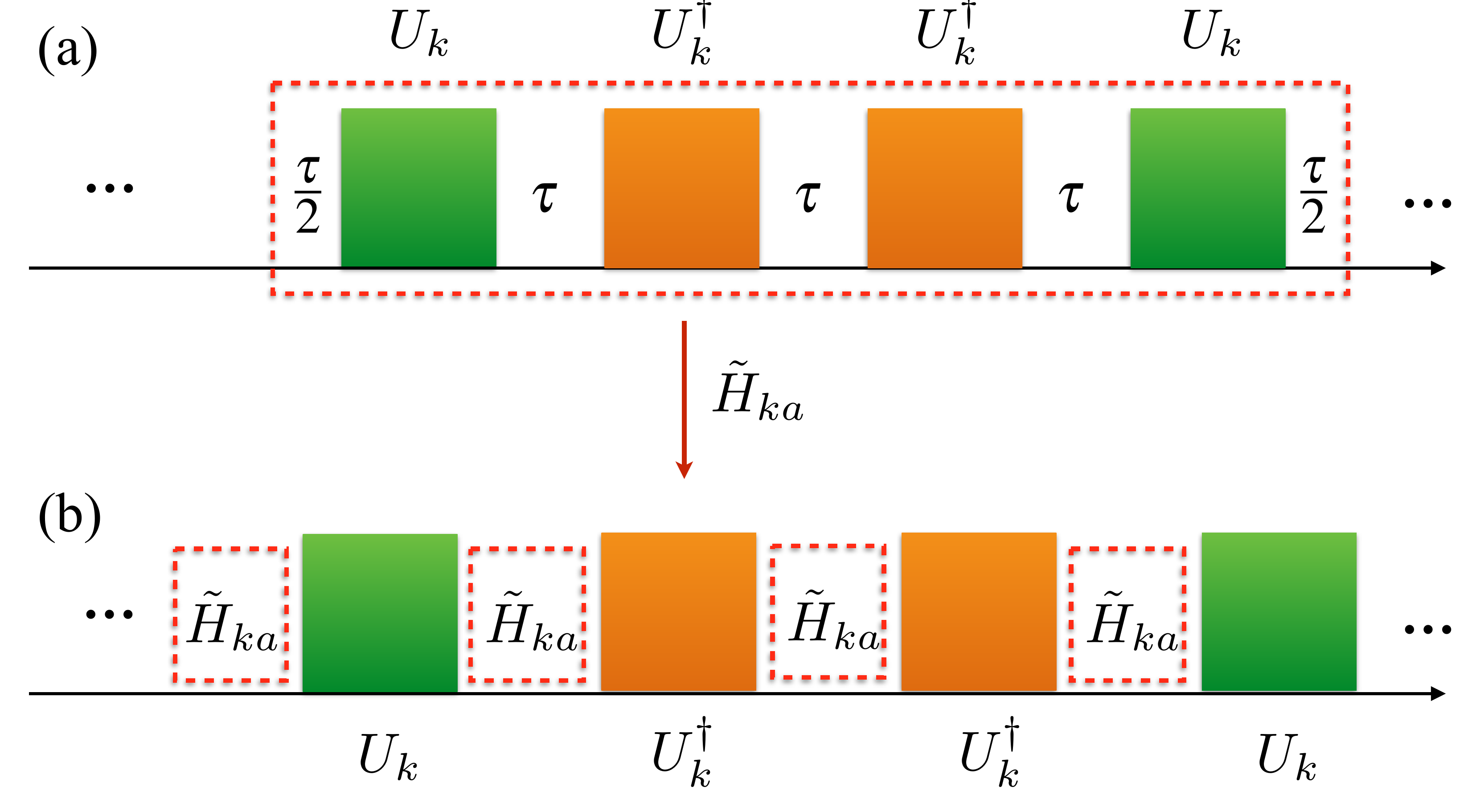}
\end{center}
\caption{(Color online) The dynamical decoupling scheme for the engineering of parameter-dependent
effective interaction Hamiltonian. {\bf (a)} The first order decoupling cycle consists of four
decoupling gates equidistant in time.  The dynamical decoupling gate $U_k=\mbox{diag}\{1,e^{i\theta(b)}\}$ with $\theta(b)\approx \pi$ is realized by the free evolution of the $k$-th sensing system for time $\tau_d={\pi/2\omega(b_0,h)}$, and
$U_k^{\dagger}=\sigma_{y}^{(1)} U_k \sigma_{y}^{(1)}$ is realized by the same free evolution as
sandwiched by $\hat{y}$-$\pi$ pulse $\sigma_y^{(k)}$. The effective Hamiltonian for such a cycle
(as indicated in the dashed block) is given by $\tilde{H}_{ka}$. {\bf (b)} The second order decoupling
cycle is constructed by the concatenating strategy from four evolution cycles (blocks) by the first
order decoupling.}
\label{fig:decoupling}
\end{figure}
\begin{figure*}
\begin{center}
\begin{minipage}{20cm}
\hspace{-2cm}
\hspace{0cm}
\includegraphics[width=8cm]{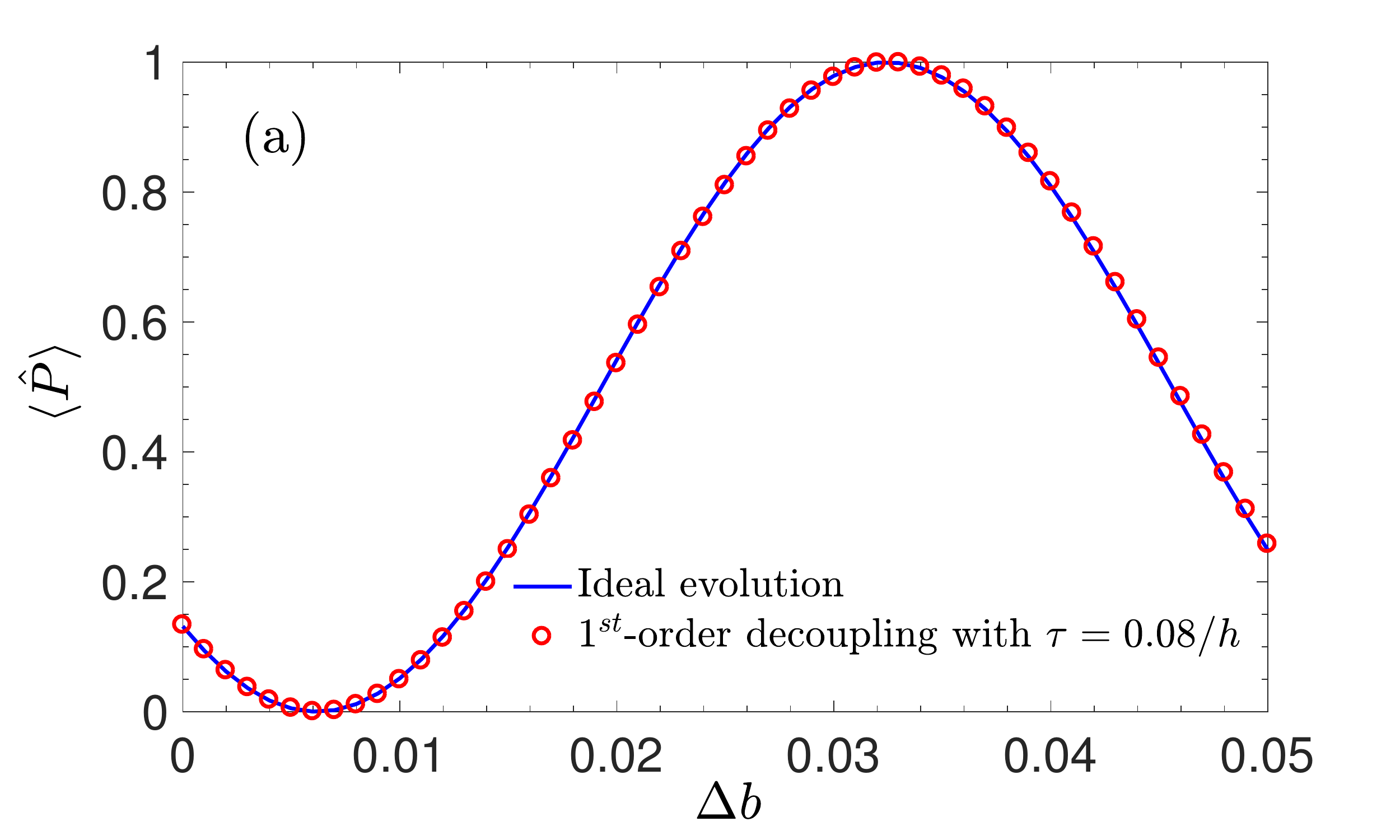}
\hspace{0cm}
\includegraphics[width=8cm]{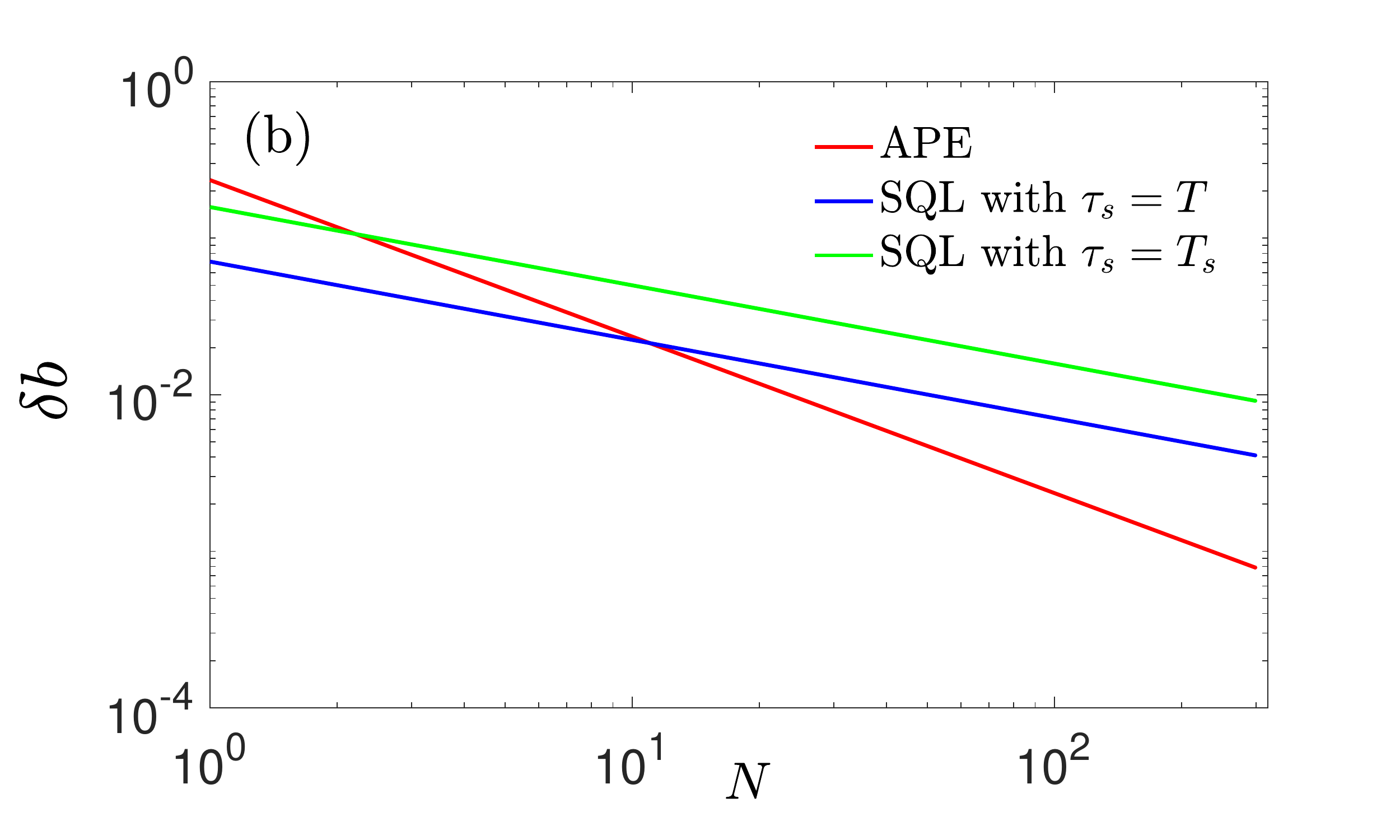}
\end{minipage}
\end{center}
\caption{(Color online) Performance of adiabatic quantum parameter
estimation. {\bf (a)}  The Ramsey signal $\langle \hat{P} \rangle$ of the probe system as a
function of the unknown parameter $\Delta b$ with the interrogation time $T_s=40/h$. The plot shows the result using first order dynamical decoupling with time interval between decoupling gates $\tau=0.08/h$ as compared with
the signal from the ideal interaction Hamiltonian $H_{k\rightarrow a}$. The system parameters
are $b_0=0.02h$, $\gamma=1.5h$, and the decoupling gate time is $\tau_d=\pi/2\sqrt{h^2+b_0^2}$.
{\bf (b)} The achievable measurement sensitivity of adiabatic quantum parameter estimation as the function of the number of sensing systems $N$ (red). We use first order dynamical decoupling with time interval between decoupling gates $\tau=0.4/h$. The interrogation time is $T_s=40/h$. The other system parameters are the same as {\bf (a)}. For comparison, we plot the standard quantum limited (SQL) sensitivity $1/\sqrt{N \tau_s}$ with the interrogation time $\tau_s$. }
\label{fig:efficiency}
\end{figure*}

To enhance the sensitivity and achieve Heisenberg limit scaling, we introduce a probe system interacting with these $N$ adiabatically
prepared systems via the Hamiltonian
\begin{equation}
    H_{ka}=-\gamma_k \sigma_z^{(k)}\otimes \sigma_z^{(a)},
\end{equation}
where $\gamma_k$ is the interaction strength that is independent on the parameter $b$. By rewriting the above
Hamiltonian in the basis of the $k$-th system $\{\ket{G_{k}},\ket{E_{k}}\}$, we obtain
\begin{equation}
    H_{ka}=\gamma_k \cos\theta \tilde{\sigma}_z^{(k)}\otimes \sigma_z^{(a)}
    +\gamma_k \sin \theta \tilde{\sigma}_x^{(k)}\otimes \sigma_z^{(a)}, \label{eq:H12}
\end{equation}
where the effective Pauli operators are defined as $\tilde{\sigma}_z^{(k)} = \bla{ \ketbra{G_{k}}{G_{k}} - \ketbra{E_{k}}{E_{k}}}$,
and $\tilde{\sigma}_x^{(k)}=\bla{ \ketbra{G_{k}}{E_{k}}+\ketbra{E_{k}}{G_{k}}}$. The first term
$H_{k\rightarrow a}=\gamma_k \cos\theta \tilde{\sigma}_z^{(k)}\otimes \sigma_z^{(a)} $ in the above
Hamiltonian represents a parameter-dependent effective field acting on the auxiliary system, which
depends on the state of the $k$-th system (namely its ground state $\ket{G_k}$ or excited state
$\ket{E_k}$). Thus, once the sensing systems are in the ground state of their local Hamiltonian,
the effective Hamiltonian for the probe system becomes
\begin{equation}
    H_a^{(E)}(t)=\sum_k \gamma_k \cos\theta  \sigma_z^{(a)} . \label{eq:H2E}
\end{equation}
The second term $H_{ka}^{u}=\gamma_k \sin \theta \tilde{\sigma}_x^{(1)}\otimes \sigma_z^{(2)}$ in the
Hamiltonian Eq.(\ref{eq:H12}), however would lead to the transition of the $k$-th system between its
ground state and excited state and would lead to deviations from the Hamiltonian in Eq.(\ref{eq:H2E})
by introducing effective dephasing \cite{SI}.

In order to eliminate such an unwanted effect, we devise a dynamical decoupling strategy
and engineer an effective Hamiltonian that eliminates the effect of the second term
in Eq.(\ref{eq:H12}). We assume that the interaction described by Eq.(\ref{eq:H12})
is switched off during the realization of the decoupling gate, which is achievable in a wide
variety of physical system. The Hamiltonian of the $k$-th sensing system after the adiabatic preparation
is $H_k(T_a) =\omega(b,h) \tilde{\sigma}_z^{(k)}$. As in quantum parameter estimation, the
parameter is $b=b_0+\Delta b$, the value of $\Delta b$ is the quantity that we want to estimate
precisely. The free evolution for time $\tau_d={\pi/2\omega(b_0,h)}$ leads to an unitary
transformation $U_k=\mbox{diag}\{1,e^{i\theta(b_0)}\}$, in the basis of $\{\ket{G_k},\ket{E_k}\}$
where $\theta(b_0)= \pi+\delta_z$, with the decoupling gate error $\delta_z \approx \pi
\bla{ {2b_0\Delta b} +{\Delta b^2} }/2\bla{h^2+b_0^2}$ \cite{SI}.
Such a unitary transformation $U_k(x)$ sandwiched by a $\hat{y}$ rotation $\sigma_{y}^{(k)}$
gives us $\sigma_{y}^{(k)} U_k (b) \sigma_{y}^{(k)} =U_k^{\dagger}(b)$. Up to the second order
of the time interval $\tau$ between decoupling gates, according to the Baker-Campbell-Hausdorff
relation, the effective interaction Hamiltonian under the dynamical decoupling sequence, see
Fig.\ref{fig:decoupling}(a), can be written as \cite{SI}
\begin{equation}
    \tilde{H}_{ka}\approx \gamma_k \cos\theta \tilde{\sigma}_z^{(k)}\otimes \sigma_z^{(a)}
    + \delta_x \tilde{\sigma}_x^{(k)}\otimes \sigma_z^{(a)}, \label{eq:amp-Hamiltonian}
\end{equation}
where we denote the quantity $\delta_x=\bla{a_x^2+a_y^2}^{1/2}$ with $a_x=\gamma_k\sin\theta\sin^2\bla{ {\delta_z}/{2}}$, $a_y=-\frac{1}{2}\gamma_k^2\tau\cos\theta\sin\theta \cos^2\bla{ {\delta_z}/{2}} $. It can be seen
that the undesirable second term in Eq.(\ref{eq:amp-Hamiltonian}) arises from both the inaccuracy
of dynamical decoupling protocol and the second order correction to the effective Hamiltonian.
The influence of the former can be further suppressed by using a higher order dynamical decoupling
sequences \cite{SI} which come at the price of a slightly increased time cost due to an
increased number of decoupling gates, see Fig.\ref{fig:decoupling}(b). In our procedure, the
number of decoupling gates increases by a factor of $5/4$ which represents a moderate time cost.
Therefore, our dynamical decoupling procedure can engineer the effective interaction
$\tilde{H}_a = \sum_k H_{k\rightarrow a} = \sum_k \gamma_k \cos\theta \tilde{\sigma}_z^{(k)}\otimes
\sigma_z^{(a)}$ as required for adiabatic parameter amplification. The total effective field acting
on the probe system is $b_a=b\sum_k \gamma_k/\omega(b,h)$. Without loss of generality, hereafter
we assume that $\gamma_k=\gamma$, and the field $b$ is amplified by a factor $\eta=N \gamma/\omega(b,h)$
which scales linearly with the number of sensing systems $N$.

{\it Analysis of achievable sensitivity.---} Once the $N$ sensing systems are
prepared into their local ground state $\ket{G_k(b,h)}$, we use the probe system to measure the
parameter information encoded in $\ket{G_k(b,h)}$ via the above engineered parameter-dependent effective
interaction Hamiltonian. In order to estimate the parameter, we apply a Ramsey sequence, namely we
initially prepare the probe system into a superposition state $\ket{\psi_a(0)} = \frac{1}{\sqrt{2}}
\bla{\ket{0_a}+\ket{1_a}}$ by applying a $\pi/2$ pulse to the state $\ket{0_a}$, after the interrogation
time $T_s$, the state evolves to $\ket{\psi_a(T_s)} = \frac{1}{\sqrt{2}}\bla{\ket{0_a} + e^{i b_a T_s}\ket{1_a}}$.
To extract the information about the unknown parameter, we apply a $\pi/2$ pulse and measure the observable
$\hat{P}=\ketbra{0_a}{0_a}$ of the probe system, see Fig.\ref{fig:efficiency}(a).

The total time $T$ for one experiment run includes the adiabatic ground state preparation time $T_a$
for the $N$ systems, and the actually time cost $\tilde{T}_s$ (accounting for the required time for the
realization of the dynamical decoupling gates) for an effective interrogation time $T_s$. The time
required to maintain the adiabatic condition during the state preparation $T_a=\bla{4b \epsilon_a}^{-1}
\left[A/\sqrt{1+A^2}-c/\sqrt{c^2+1} \right]$ with the ratio $A = \lambda_k(0)/b $ and $c =h /b $, is
obtained from the adiabatic condition ${\vert \bra{e(t)}\dot{H}_{k}(t)\ket{g(t)}\vert }/{\vert E_e -
E_g\vert ^2 }\equiv \epsilon_a \ll 1$, where $\ket{g_k(t)}$ and $\ket{e_k(t)}$ are the instantaneous
eigenstates of the Hamiltonian $H_k(t)$, and $E_{g,e}=\pm \blb{\lambda_k(t)^2+b ^2}^{1/2}$ are the
corresponding eigenenergies. The small quantity
$ \epsilon_a$ characterizes how well the adiabatic condition is fulfilled. The adiabatic ground state
fidelity is $1-\delta \approx 1-e^{-T_a\Delta/v}$ \cite{Landau32,Zener32}, where $v$ is the relative
slope of the energy levels and $\Delta$ is the minimal energy gap between the ground and excited state.
An effective interrogation time $T_s$ will require a realization time $\tilde{T}_s=T_s\bla{1+\tau_d/\tau}$,
where $\tau$ is the time interval between dynamical decoupling gates. Therefore, the total time for one
experiment run is $T=T_a+T_s\bla{1+\tau_d/\tau}$. The achievable shot-noise limited sensitivity for the
estimation of the parameter $b$ is given by $\delta b=\sqrt{\langle \Delta^2 \hat{P} \rangle}/
\bla{\frac{\partial}{\partial b}{\langle \hat{P} \rangle} \sqrt{1/T} }$. In the absence of noise and
under the assumption that $h\gg b$, we obtain the achievable sensitivity as \cite{SI}
\begin{equation}
    \delta b \approx \frac{1}{N}\bla{\frac{h} {\gamma}} \frac{\sqrt{T}}{{T_s}} =
    \frac{1}{N}\bla{\frac{h} {\gamma}} \frac{\sqrt{T_a+T_s\bla{1+\tau_d/\tau}}}{{T_s}}
\end{equation}
It can be seen that the sensitivity reaches a Heisenberg limit scaling $ \delta b  \sim 1/N$, although
the parameter amplification is accompanied by the extra time cost required for the realization of dynamical
decoupling gates leading to an N-independent additional factor. As compared with standard quantum limit $\bla{\delta b}_{SQL}=1/\sqrt{N T}$, the enhancement factor is $\delta b / \bla{\delta b}_{SQL} =\bla{h/\sqrt{N}\gamma }\bla{T/T_s}$, which is more pronounced as the value of $N$ increases, see Fig.\ref{fig:efficiency}(b). The additional advantage of the present proposal comes from the suppression of the effect of non-Markovian noise. In a standard quantum limited scheme, the interrogation time will be limited to $T_2^*$. While in the present scheme, the interrogation time is limited by the lifetime of the sensing systems (as protected by the energy gap) and the coherence time of the probe system. The probe system only needs to couple with the sensing system and may decouple from noise, thus the interrogation time (which may be further extended using decoherence free subspace \cite{Dorner12,Kotler14}) would be much longer than $T_2^*$ of the sensing system.

The implementation of adiabatic quantum parameter estimation requires the capability of tuning driving
field and engineering Hamiltonian, which is feasible in current state-of-art experiment setups. For
example, we note that both analog adiabatic quantum simulation and digital quantum simulation, demonstrating
these experimental capabilities, has achieved considerable progress in several types of physical systems. In
particular, the techniques for the engineering of various spin-spin interactions and the coherent manipulation
of ion spin state with a high fidelity have been very well developed for trapped ions \cite{Blatt12,Boh16}.

{\it Effect of imperfection and entanglement.---} We provide a detailed analysis of the two main sources of imperfect coherent control in the above protocol. In the adiabatic ground state preparation, due to the finite
preparation time, the final state of the $k$-th system $\psi_k(T)=\sqrt{1-\delta}\ket{G_k}+\sqrt{\delta }\ket{E_k}$
is not the exact ground state and results in a ground state fidelity $1-\delta$ ($\delta\ll 1$). The
excitation in the sensing systems leads to an effective dephasing in the probe system. The best
achievable sensitivity under the influence of the imperfect ground state preparation,choosing the interrogation time $T_s$ such that ${(\gamma/h) bT_s}=k\pi$ with $k\in \mbox{odd}$,
is found to be the same as the ideal case, namely
 $   \delta b \approx \frac{1}{N}\bla{ \frac{h} {\gamma} } \frac{\sqrt{T}}{{T_s}}$ \cite{SI}.
The second imperfection lies in the high order corrections to the engineered Hamiltonian that we
obtained via the the dynamical decoupling procedure shown in Eq.(\ref{eq:amp-Hamiltonian}). The transformation
of the probe system due to the interaction of the $k$-th system is described by a completely positive
map $\mathcal{M}(\rho_a) = (1-\delta')U_z \rho_a U_z^{\dagger}+\delta' \sigma_z \rho_a \sigma_z$, where
$U_z=\exp{(-iT_s \gamma\cos\theta {\sigma}_z)}$ is the ideal evolution, and $\delta'\lesssim \delta_x^2/
\bla{\gamma^2 \cos^2\theta}$ \cite{SI}. Therefore, the final state of the probe system accounting for the high order
corrections is $\rho_a(T_s)=\otimes_j \mathcal{M}_j(\rho_a(0))$, which leads to the best achievable sensitivity as $   \delta b \approx \frac{1}{N}\bla{ \frac{h} {\gamma} } \frac{\sqrt{T}}{{T_s}}$ under the same condition as ${(\gamma/h) bT_s}=k\pi$ with $k\in \mbox{odd}$. It can be seen that the two main sources of imperfect control generally leads to a reduction of the achievable sensitivity, but the reduction can be compensated for by choosing an appropriate interrogation time, and thus will not  affect the achievable sensitivity.

Quantum entanglement appears unintentionally in the interrogation step due to the imperfect ground state preparation and due to
the corrections to the ideal effective Hamiltonian. In standard quantum metrology, the role of entanglement
in the initially prepared state has been carefully studied \cite{Cze15,Aug16}. In the present scenario however,
entanglement does not seem to play the role of quantum resource for parameter estimation. In general, the appearance of entanglement is accompanied with the reduction in the achievable sensitivity \cite{SI}. We choose the most suitable interrogation time $T_s$ that achieves the best sensitivity, entanglement instead disappears.

{\it Conclusion.---} In summary, we have proposed a paradigm of adiabatic quantum parameter estimation to achieve high measurement sensitivity in the presence of noise. In particular, the present proposal benefits from the energy gap protection and provides an efficient strategy for estimating local Hamiltonians against parallel phase noise. The techniques can be readily realized with the current state-of-art quantum technology, for example using trapped ions. We demonstrate that it allows to achieve Heisenberg limited measurement sensitivity without relying on quantum entanglement as a resource.  Our proposal thus provides a platform to help elucidate the fundamental role of quantum entanglement in quantum metrology.

{\it Acknowledgements.---} We thank Prof. Ren-Bao Liu for the fruitful discussion. J.-M.C is supported by the National Natural Science Foundation of China (Grant No.11574103), the National Young 1000 Talents Plan. M.B.P is supported by an Alexander von Humboldt Professorship, and ERC Synergy grant and the EU project DIADEMS, EQUAM and QUCHIP.

\end{document}